# Dynamics of chirped Airy pulse in a dispersive medium with higher-order nonlinearity

ANKIT PUROHIT, DEEPENDRA SINGH GAUR, AND AKHILESH KUMAR MISHRA*

*Department of Physics, Indian Institute of Technology Roorkee, Roorkee-247667, Uttarakhand, India*
*\*akhilesh.mishra@ph.iitr.ac.in*

**Abstract:** Chirp can control the dynamics of the Airy pulse, making it an essential factor in pulse manipulation. Finite energy chirped Airy pulse (FECAP) has potential applications in underwater optical communication and imaging. Hence, it's critical to study the propagation of FECAP. We present a numerical investigation of the propagation dynamics of a FECAP in a dispersive, and highly nonlinear medium. The nonlinearity under study includes self-phase modulation (SPM), self-steepening (SS), as well as intra-pulse Raman scattering (IRS) terms. We have observed soliton shedding and the chirp parameter is demonstrated to have a considerable impact on the pulse dynamics. In particular, the emergent soliton does not propagate in a straight path instead, depending on the sign of the chirp parameter, it delays or advances in the time. Furthermore, it has been established that the chirp can be employed as an alternate control parameter for the spectral manipulation. The results of our study may have implications in supercontinuum generation and for producing tunable sources.

## 1. Introduction

In 1979, Berry and Balaz introduced Airy wave packet in the context of quantum mechanics [1]. The intriguing properties of these wave packets were being diffraction-less and self-acceleration [2]. Since these wave packets contain an infinite amount of energy, they remain unaffected to diffraction. On the other hand, free acceleration means that these waves follow a nonlinear (parabolic, in particular) trajectory in the absence of any external potential. The study of the accelerating wave starts by understanding the fact that the acceleration in such waves is not manifested by the centroid (which moves in a straight line to preserve the transverse momentum), but it is solely due to the accelerating field structure [2]. The parabolic trajectory of the Airy beam can be manipulated by imposing a linear phase at the input [3]. Due to infinite extension, these wave packets couldn't be realized experimentally. In 2007, Siviloglou and Christodoulides proposed the concept of finite energy Airy beams (FEABs) theoretically and later demonstrated these beams in lab experimentally [4,5]. The diffraction free ideal Airy beam on passing through a finite aperture becomes truncated and hence produces diffraction. A truncation factor can control the extent of this diffraction. Thus, FEAB is quasi non-diffractive beam. Another interesting characteristic of the Airy beam is that it can be observed even in one dimension (1D). However, all other non-diffracting beams such as Bessel beam, Mathieu beam can't be anticipated in 1D [2]. FEABs have found applications in many fields, such as optical coherence tomography, curved plasma guidance [6], optical trapping and manipulations [7-9], linear light bullet generation [10,11] and high-resolution microscopy [12,13] among others. In addition, there are reports on Airy vortices with angular momentum, which have application in communication [14-18].

An ideal Airy beam is shape preserving accelerating solution of paraxial equation of diffraction in space. Owing to the mathematical equivalence between "Paraxial equation of diffraction in space" and "temporal envelope dispersion equation in time", temporal Airy pulses were introduced. The ideal Airy pulse is a solution of temporal dispersion equation in time. Finite energy Airy pulse (FEAP) also exhibits remarkable properties of being quasi non-dispersive, self-bending, and self-acceleration as those observed in the FEAB [19]. Physically, the interpretation of temporal acceleration is quite different from the spatial acceleration. The spatial acceleration term implies that its trajectory bends along a parabolic trajectory in space, however, temporal acceleration intends to change in velocity of the intensity peak of the pulse that manifests as self-acceleration or self-deceleration depending on whether its side lobes are at leading or trailing edge of the main peak [19]. A lot of investigations have already been done on the propagation of both un-chirped [20,21] and chirped Airy pulse [22,23] in dispersive media. In nonlinear domain, Airy pulses have shown soliton shedding [24, 25], modulation instability [26], and filamentation [27]. An initial chirp can be employed in a nonlinear dispersive medium to impact laser self-focusing considerably [28]. The chirp-based symmetric Airy pulse would be employed to generate a focal point in the propagation axis [29]. In addition, asymmetric Airy pulses can also generate and regulate multicolor Raman soliton with improved tunability [30]. FECAP has potential applications in underwater optical communication and imaging, was recently highlighted. [31]. These fascinating and unusual properties, particularly in nonlinear media, have revealed enormous potential towards practical applications [32]. These observations demand further exploration of the dynamics of such pulses with higher order nonlinearity. To the best of our knowledge, none has previously reported the evolution of chirped ultra-short Airy pulse with higher order nonlinear effects such as SS, IRS etc. Hence it becomes essential to analyze the propagation dynamics of these pulses in such a medium. This study may find applications in realizing pulse compression, supercontinuum generation, and filamentation.

## 2. Numerical model

We model the propagation of a pulse in a dispersive and nonlinear medium using the nonlinear Schrödinger equation (NLSE) [24]-

$$\frac{\partial A}{\partial Z} + \frac{i\beta_2}{2}\frac{\partial^2 A}{\partial \tau^2} = i\gamma_1 |A|^2 A - \frac{\gamma_1}{\omega_0}\frac{\partial}{\partial \tau}(|A|^2 A) - i\gamma_1 \tau_r \frac{\partial}{\partial \tau}(|A^2|A) \tag{1}$$

where $A$ is the envelope of the slowly varying optical field, $\beta_2$ the second-order dispersion coefficient, $Z$ the propagation direction, $\gamma_1$ is the cubic nonlinearity coefficient, $\omega_0$ is the central frequency, $\tau_r$ is the decay time of the Raman gain, and $\tau$ is the time in pulse frame of reference. $\tau$ is related to time t in lab frame via the relation $\tau = t - Z/v_g$, where $v_g$ is the group velocity of the pulse. In numerical modelling, it is customary to write NLSE in dimensionless form-

$$\frac{\partial U}{\partial z} + i\frac{sgn(\beta_2)}{2}\frac{\partial^2 U}{\partial T^2} = N^2\left(i|U|^2 U - s\frac{\partial}{\partial T}(|U|^2 U)\right) - iN^2 T_r \frac{\partial}{\partial T}(|U^2|U) \tag{2}$$

To get the Eq. (2) from Eq. (1), following transformations have been performed

$$z = \frac{Z}{L_{D2}}, T = \frac{\tau}{\tau_0}, U = \frac{A}{A_0}, T_r = \frac{\tau_r}{\tau_0}.$$

Here $L_{D2} = \tau_0^2/|\beta_2|$ is second-order dispersion length, $L_{NL} = 1/\gamma_1|A_0|^2$ is nonlinear length, $N^2 = L_{D2}/L_{NL}$, represents the strength of Kerr nonlinearity, $s = 1/\omega_0\tau_0$ is the self-steepening parameter; $z, T, U, T_r$ respectively represent dimensionless distance, dimensionless time, dimensionless envelope, and dimensionless Raman gain time. $sgn(\beta_2) > 0$ corresponds to normal dispersion while $sgn(\beta_2) < 0$ correspond to the anomalous dispersion regime.

The FECAP at input is expressed as follows [22]-

$$U(T, z = 0) = \sqrt{f(a)}Ai(T)exp(a\tau)exp(-iCT^2) \tag{3}$$

where $C$ is chirp parameter and 'a' is truncation factor. We would like to note that throughout the manuscript, unless otherwise specified, chirp indicates linear chirp only. In our simulation, 'a' is taken to be 0.2. $Ai$ stands for airy function. Usually, $f(a)$ is chosen such that the input pulse peak intensity becomes unity.

### 3. Results and Discussions

We solve Eq. (2) for chirped Airy pulse by employing well known split-step Fourier transform method in anomalous group velocity dispersion (GVD) regime for focusing nonlinearity [33]. The exact cancellation of the induced negative chirp of Anomalous GVD with the positive chirp induced by focusing nonlinearity (self-phase modulation) generates temporal soliton [34]. The equilibrium condition for soliton generation reads as $L_{D2} = L_{NL}$ but there are cases of perturbed soliton (that is when either dispersion or nonlinear effect dominates slightly or when pulse even at the input is chirped) propagation; where the maximum intensity of pulse oscillates periodically [25, 34]. In the following sections our main focus is to investigate the effect of chirp on soliton formation (soliton shedding) and on the resultant pulse spectrum. We also study the roles of self-steepening and intra-pulse Raman scattering on chirped Airy pulse evolution.

#### 3.1 $T_r = 0$ and $s = 0$

In this section, we focus our attention on the soliton formation in the anomalous-Kerr medium for differently chirped input pulses. We assume that the Airy pulse has enough input power (i.e., N≥1) to shed soliton. Propagation of Airy pulse in anomalous-Kerr medium leads to the formation of soliton pulse while a fraction of the side lobe power follows the usual parabolic trajectory and self-heals itself to maintain its Airy nature. Previous studies have shown that a large amount of energy is spread in side lobes at a small truncation factor (a = 0.1) due to the feature of acceleration (self-healing) of the Airy pulse [24]. Therefore, we choose a large truncation factor (a = 0.2) so that most of the power be concentrated in main lobe of the pulse. To investigate the effect of chirp on FECAP, we fixed the self-phase modulation (SPM) term such that N=1.5 and truncation factor (a = 0.2) and vary the chirp in the input Airy pulse. Fig. 1(a, d) shows the temporal and spectral evolution of un-chirped (transform limited) Airy pulse with propagation. For initially un-chirped Airy pulse, a part of pulse of relatively high power gets separated from the main Airy pulse and propagates as a solitonic pulse. This pulse appears to propagate a little early in the time and move in a straight-line path. However, for positive chirp (C =0.3) input pulse, quasi soliton forms. This quasi soliton shows a large temporal delay in time and follows a slanted linear path as shown in Fig. 1(b, e). In addition, small amplitude dispersive wave appears at the trailing edge and subsequently gradually depart from main soliton with the propagation [24]. For initially negative chirp (C =-0.3) pulse, quasi soliton appears advanced in time and follows oppositely slanted straight trajectory. Here, the small amplitude dispersive wave appears at the leading edge [24]. In all of the above cases, as stated earlier the remaining pulse heals itself and maintains its Airy behaviour. The solitons thus formed (or shed) propagate in dispersive background and the interplay of dispersion and nonlinearity induces an oscillatory effect as visible in Fig. 1(a, b and c). Fig. 1(g), shows the relative temporal positions of these three differently chirped input Airy pulses after a propagation of z=25. On the other hand, if we include nonlinear chirp (such as quadratic) in input pulse, the pulse evolution becomes more complex [35,36].

The spectral domain evolutions of the above discussed cases are shown in Fig. 1 (d, e and f). For unchirped (C=0) Airy pulse, the spectrum is Gaussian in shape at input (z=0). With propagation the spectrum gradually changes over to a typical oscillatory and multipeak asymmetric structure (see Fig. 1(d)). The oscillations on spectrum can be understood by the fact that at different time points, the pulse may have the same instantaneous frequency but different phase that can interfere constructively or destructively as per their relative phase differences [24, 34]. For the positive chirp Airy pulse (C = 0.3) as shown in Fig. 1(e), the spectrum has Airy distribution even at the input (z=0), while at the output (z = 25) we see red-shifted frequency component. In particular, at the output, Airy distribution like oscillatory structure appear at the edge of red-shifted spectra. In contrast, in case negatively chirped (C= -0.3) Airy pulses as shown in Fig. 1(f), at input (z=0) the spectrum is mirror image of that of the positively chirped Airy pulse i.e. the spectrum is blue-shifted. Even on propagation the spectrum almost mimics the mirror image of the spectrum of the positively chirped Airy pulse. On the same note, Airy distribution like oscillatory structure now appears at the edge of the blue-shifted spectra. Fig. 1(h) shows the relative spectral positioning of the spectrum of above mention pulses.

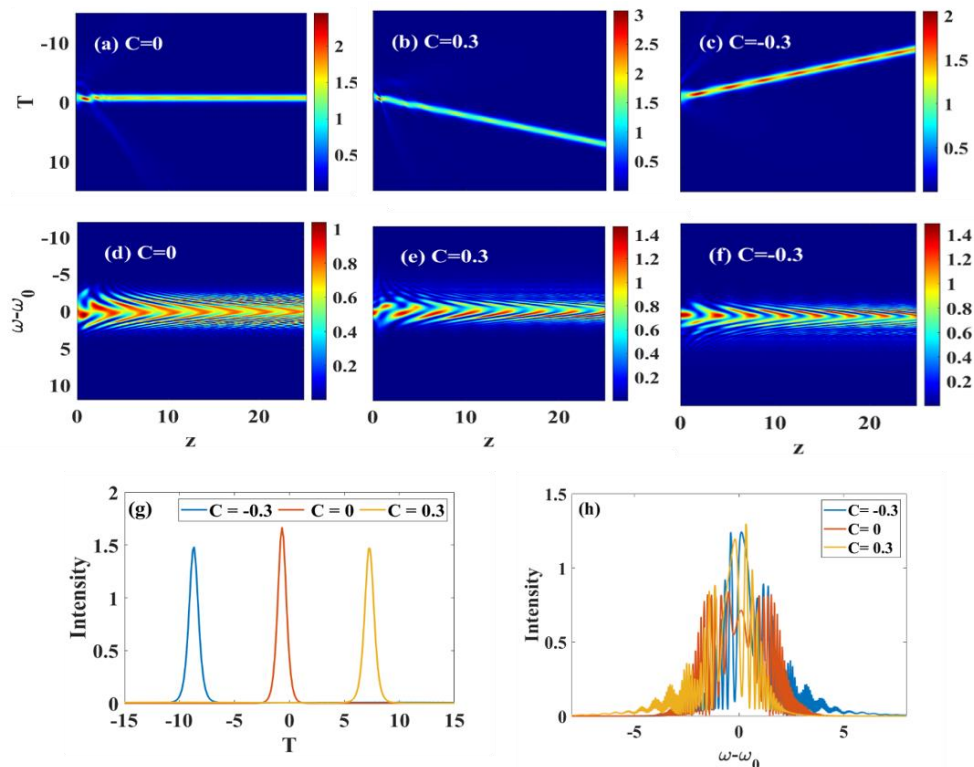

Fig. 1. Temporal and Spectral evolution of FECAP in anomalous dispersion and focusing nonlinearity (N = 1.5): Contour maps of propagation dynamics of Airy pulse with (a), (d) C =0, (b), (e) C = 0.3, and (c), (f) C = -0.3; (g) Temporal intensity, and (h) Spectral intensity of output pulse at z = 25.

The initial studies show that the SPM effect tends to dominate with the increase in the input Airy pulse peak power (i.e. N) that results in the periodic variation in the maximum intensity (MI) of the Airy pulse [25]. To account for the effect of chirp, we have plotted MI with propagation distance z in Fig. 2 (a) for different values of chirp parameter while keeping N (=1.5) fixed. Pulse's MI shows a periodic variation with propagation. For unchirped pulses (C=0), this oscillation is relatively small. The MI begins to oscillate with a larger amplitude with the increasing magnitude of chirp (irrespective of the numerical sign of chirp). The pulse width of soliton is directly related to the MI. The width of solitons is inversely associated with the amplitude of solitons [34]. Therefore, the width of a soliton also varies periodically.

Since temporal position of the output pulse also depend on the chirp, we have plotted the evolution of temporal position of maximum intensity (PMI) with distance for different input chirp values in Fig. 2 (b). For un-chirped Airy pulse (C=0), the optical soliton appears a little earlier and afterward propagates in straight line path. The soliton's trajectory changes dramatically for chirped Airy pulse. Positive chirp delays the soliton and the delay proportionate directly to the input chirp values. On the contrary, the negative chirp results time advanced soliton and this advancement is also directly proportional to the input chirp parameter. Thus delay/advancement of optical solitons are linearly proportional to the magnitude of the chirp parameter and to the propagation distance.

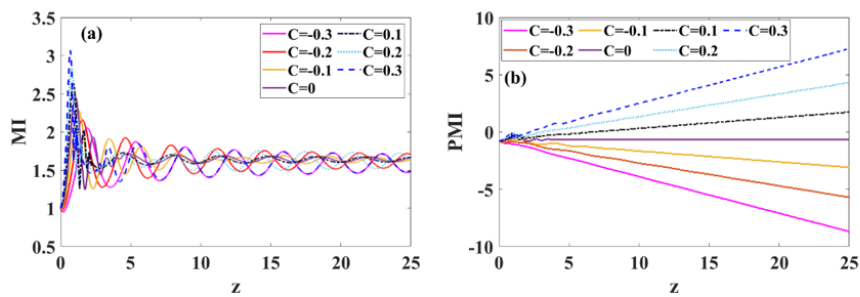

Fig. 2. (a) MI and (b) PMI plotted as a function of propagation distance for differently chirped Airy pulses (N=1.5).

### 3.2 $T_r \neq 0$ and $s = 0$

Up till now, we have not taken SS and IRS induced effects into account. In this section, our primary focus is to model the evolution of chirped Airy pulse in the presence of IRS for differently chirped input pulse while neglecting SS term in Eq. (2). IRS, also named as Raman induced frequency shift (RIFS), is one of the most sensitive high-order nonlinear effect. IRS shifts the frequency spectrum toward the longer wavelength side [34]. In 1986, Mitschke and Mollenauer, for the first time, pointed out this effect [37], later, Gordon elaborated it by giving theoretical explanation [38]. Since then, the influence of the IRS on several types of pulses, such as Gaussian, super-Gaussian etc., has been investigated extensively [34]. IRS manifests itself by compressing the main lobe of pulse initially, and then bending pulse's trajectory gradually with further propagation [24]. This bending is directly proportional to the parameter N. Furthermore, a part of the energy is lost in dispersive media due to the free acceleration nature of Airy pulse that can be controlled by manipulation of the truncation factor [24]. In our analysis, we have fixed N and $T_r$ parameters at 1.5 and 0.05 respectively and varied the input chirp of the FEAP. The evolution of such a pulse is plotted in Fig. 3. As depicted in the first row of Fig. 3, in time domain the major effect of IRS is to bend the pulse trajectory and we also know that the pulse trajectory is extensively influenced by the initial chirp parameter. For C=0, initially, the interplay of dispersive, SPM, and IRS disrupts the pulse shape, but with the propagation, nonlinear effect (mainly IRS) dominates. The pulse is delayed in the time domain. For C=0.3, as shown in Fig. 3 (b), the time delay is reduced as compare to C=0 case. In contrast, for C=-0.3, the delay gets enhanced as compare to that of C=0 case (see Fig. 3(c)). Additionally, we have also noted that if we increase the $T_r$, e.g. for $T_r = 0.1$, the positively chirped pulse splits into several tiny pulses (Figure not shown). While, for negatively chirped pulse, for larger $T_r$, the delay time further increases and the resultant pulse consists of high amplitude and short pulse width features. The effects of IRS on differently chirped pulses are evident in Fig. 3(g). A comparison between Fig. 1(g) and Fig. 3(g) also establishes the role of IRS in introducing time shift.

A recent study on FEAP reveals that N parameter's high values stimulate the smoother and greater red-shifted (long wavelength) spectra, but the blue-shifted spectra are almost the same [24]. To disclose the effect of chirp on pulse spectrum, we have also shown the corresponding pulse spectra for the cases discussed above (that is C = 0, 0.3, -0.3), respectively in Fig. 3(d), 3(e), and 3(f). The spectrum of a positively chirped pulse develops very complex structures and shrinks with propagation. However, for C = -0.3, the pulse spectra get broadened (see Fig. (3) h) as compare to those of the other two cases.

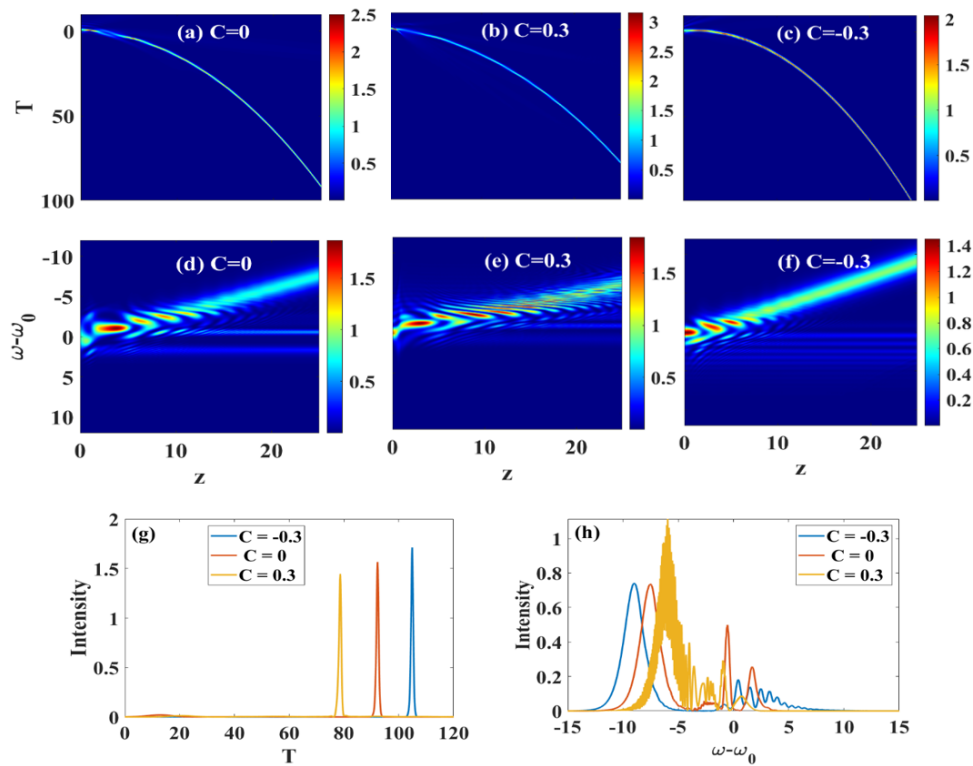

Fig. 3. Temporal and Spectral evolution of FECAP in anomalous dispersion, focussing nonlinearity (N = 1.5), and $T_r$ =0.05 medium: Contour maps at (a), (d) C =0, (b), (e) C = 0.3, and (c), (f) C = -0.3; output pulse (g) Temporal intensity, and (h) Spectral intensity at z = 25.

Variation of MI with propagation is shown in Fig. 4(a). For negatively chirped pulse, MI initially varies periodically, later with propagation it becomes smooth. On the other hand, for positively chirped pulse, MI shows oscillatory variations even after relatively longer distance of propagation. Therefore, the initial chirp is an essential factor in deciding the peak intensity of the FEAP.

PMI is another crucial parameter to characterize the dynamics of FECAP. As shown in Fig. 4(b), initially there is almost no change in PMI, but with further propagation, PMI changes significantly with chirp parameter. As mentioned earlier, the delay in time domain increases with initial negative chirp, but it is not always true. If we increase the chirp value further, the delay in the time domain decreases afterward. Similarly, in positive chirp, the PMI first bend less as compared to C = 0 case, further increasing the chirp leads to increase in the bend. Therefore, the initial chirp is a critical factor to manipulate the shift in the time domain with the distance.

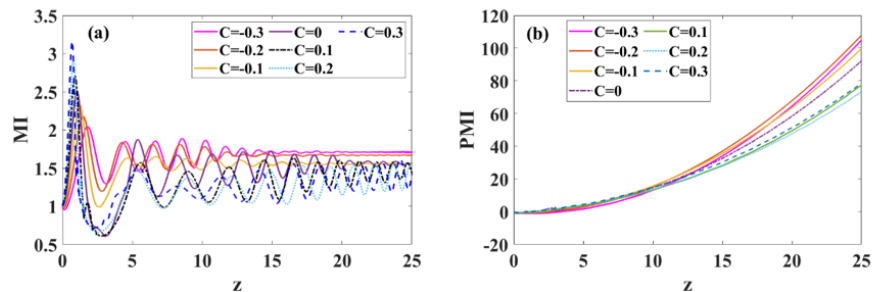

Fig. 4. (a) Maximum intensity (MI) and (b) Position of MI plotted as a function of propagation distance for different chirp values when Airy pulse propagate in a medium with anomalous dispersion, focussing Kerr nonlinearity (N=1.5) and $T_r$ =0.05 in Time domain.

### 3.3 $T_r$= 0 and s ≠ 0

In this section we neglect IRS and include SS effect in our investigation. Other simulation parameters are kept the same (N=1.5). For C = 0 and s = 0.05, Fig. 5 (a) shows delayed soliton shedding. The delay is even longer than that in Fig. 1 (a) (where SS term has not been considered). The reason behind this longer delay is steepening of the pulse towards its trailing edge, a typical of SS effect. For positively chirped pulse (C=+0.3), as is shown in Fig. 5 (b), the delay is increased in presence of SS effects. Notice that while discussing Fig. 1 (b), we found that the positive chirp delays the soliton shedding. The delay in the present case is enhanced relatively. In contrast, for negatively chirped FEAP (C = -0.3), the amount of time advancement of the soliton is less (see Fig. 5 (c)) as compare to that in Fig. 1(c). Temporal positions at the output for differently chirped pulses are shown in Fig. 5 (g). We would like to note here that the shifts in Fig. 5 (g) are far different as compared to what we observe in Fig. 3 (g). On the other hand, there is no substantial change in spectra in presence of the SS term in Fig. 5(h) when compared to the non-SS case (Fig. 1(h)). SS can produce optical shock, similar to the acoustic shock produced by sound waves, therefore the coefficient of SS terms must be chosen consciously [34].

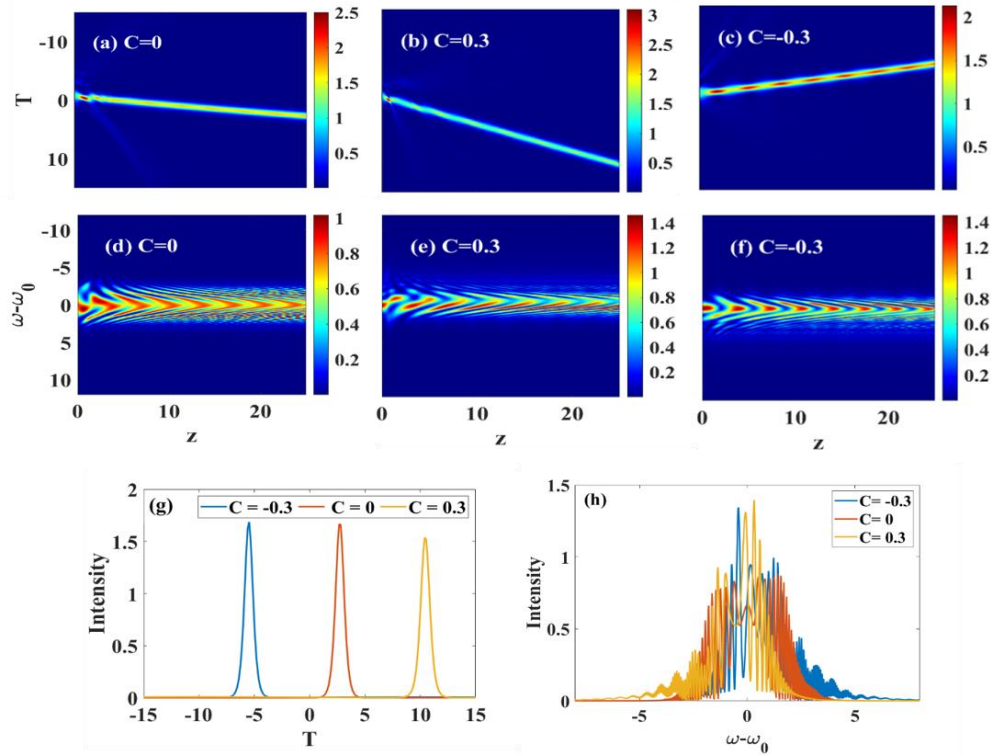

Fig. 5. Temporal and Spectral evolution of FECAP in anomalous dispersion, focusing nonlinear (N = 1.5), and s =0.05 medium: Contour maps at (a), (d) C =0, (e) C = 0.3, and (c), (f) C = -0.3; output pulse (g) Temporal intensity, and (h) Spectral intensity at z = 25.

To comprehensively understand the FECAP dynamics, we have plotted MI, for N = 1.5, s = 0.05 in the anomalous dispersion regime for different chirp values in Fig. 6 (a). On comparing Fig. 2 (a) and Fig. 6 (a) for C = 0, we found almost no distinction apart from a slight average increase with propagating in Fig. 6 (a). However, non-zero chirps do introduce differences. In Fig. 2 (a), MI for positive (C =0.3), and negative chirp (C = -0.3) are almost superimposed. In contrast, in Fig. 6 (a), the MI of negative chirp (C = -0.3) is greater than that of the MI of positive chirp (C =0.3). The corresponding PMI variation with the respective chirp values are plotted in Fig. 6 (b). Thus, SS offers an alternative to manipulate the pulse time trajectory. We would like to note that initial chirp doesn't change the pulse shape in the time domain in the absence of GVD.

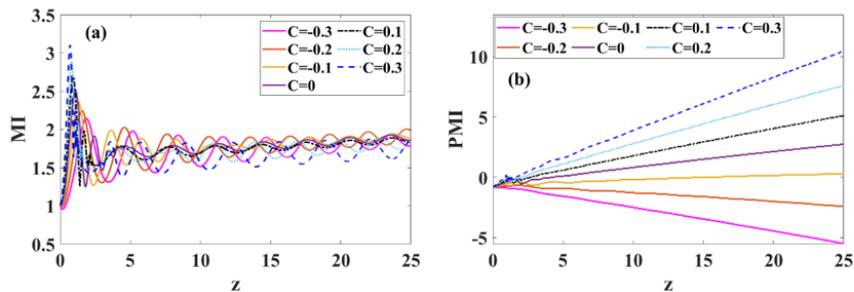

Fig. 6. (a) Maximum intensity (MI) and (b) Position of MI plotted as a function of propagation distance for different chirp values when Airy pulse propagate in a medium with anomalous dispersion, focussing Kerr nonlinearity (N=1.5) and s=0.05 in Time domain.

### 3.4 $T_r \neq 0$, $s \neq 0$

In this section, we have included all the above discussed higher-order nonlinear terms in FECAP evolution simulation. The simulation parameters are- N = 1.5, s = 0.05 and $T_r = 0.05$. For C = 0, the previous study has shown that SS effects decelerate the main lobe movement, showing the reduction in RIFS (Raman induced frequency shift) in the spectral region (compare the Figs. 3 (a), 3(d) and Figs. 7(a), 7(d)) [2, 24]. For C = 0.3, the pulse deceleration increases slightly and the spectra gets squeezed a bit when compared to C = 0 case (see Figs. 7(b) and 7(e)). On the other hand, for C = -0.3, pulse accelerates more as shown in Fig. 7(c), and the pulse spectrum broadens substantially (see Fig. 7(f) and Fig. 7(h)). Note that the spectrum in Fig. 7(h) is less broad than that in Fig. 3(h). This is because of Raman induced frequency shift. The analysis shows the nature of the chirp is a crucial parameter to manipulate the spectrum. The comparative temporal positioning is shown in Fig. 7(g). These results show that the initial chirp has the potential to produce and manipulate both time and spectral domains and may lead to the generation of supercontinuum.

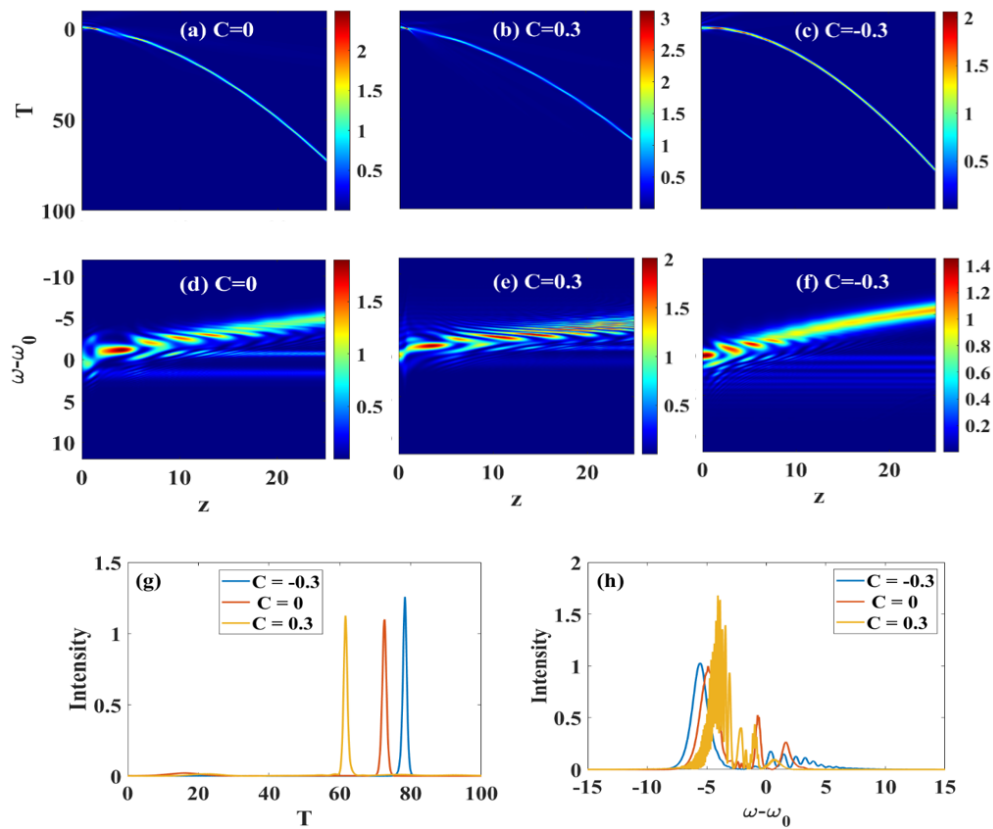

Fig. 7. Temporal and Spectral evolution of FECAP in anomalous dispersion, focussing nonlinear (N = 1.5), s = 0.05 and $T_r$ =0.05 medium: Contour maps at a), d) C = 0, b), e) C = 0.3, and c), f) C = -0.3; output pulse g) Temporal intensity, and (h) Spectral intensity at z = 25.

To understand the absolute nature of the Airy pulse's propagation dynamics, it is essential to look over the change in MI when all terms of Eq. (2) are considered. As shown in Fig. 8(a), the variation in MI with distance at different chirp. On comparing Fig 4(a) (zero SS case) with Fig. 8(a) one can conclude that with the inclusion of both SS term, the number of oscillations in MI reduces significantly. It is also apparent that FECAP propagates smoothly in negative chirp and shows undulation in positive chirp. Moreover, negative chirp has larger average MI. Therefore, we can conclude that the initial chirp is crucial in manipulating the MI of FEAP.

Change in PMI with z at different chirp values are shown in Fig. 8(b). Shifts in PMI in Fig. 4(b) and Fig. 8(b) display the broad picture of how the inclusion of the SS parameter deteriorates the IRS bending effect. Earlier, as already pointed out, bending rigorously dependents on the strength of the chirp parameters. Therefore, SS presence offers an alternate way to manipulate the pulse trajectory.

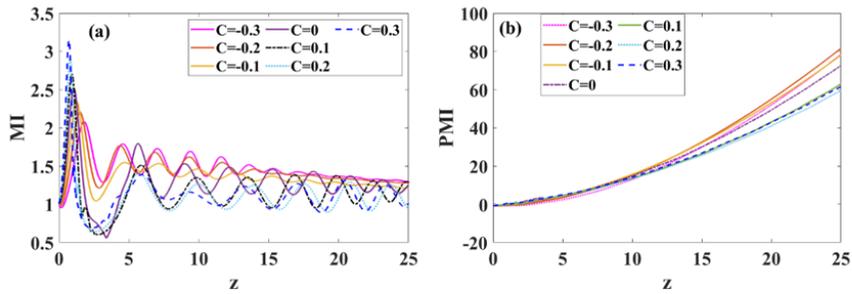

Fig. 8. (a) Maximum intensity (MI) and (b) Position of MI plotted as a function of propagation distance for different chirp values when Airy pulse propagate in a medium with anomalous dispersion, focusing Kerr nonlinearity (N=1.5), $T_r = 0.05$ and s=0.05 in Time domain.

For experimental validation of the theoretical results we must have information of the real pulse and the material conditions. To this end we would like to note that for ultrashort input airy pulse, we can choose following real parameters- pulse width $\tau_0 = 100$ fs, GVD coefficient $\beta_2 = -280\ fs^2/cm$ and power $|A_0|^2 = 1000$ W. The nonlinear refractive index for bulk silica glass given by $n_2 = 2.2 \times 10^{-20} m^2/W$ [34]. The Kerr nonlinear coefficient $\gamma_1 = 6.3\ W^{-1}Km^{-1}$ can be calculated for a given wavelength and material structure (effective core area of fiber). The normalized parameters thus obtained are - $L_{D2} = \tau_0^2/|\beta_2| = 0.3571\ m$, $L_{NL} = 1/\gamma_1|A_0|^2 = 0.1587\ m$ and $N^2 = L_{D2}/L_{NL} = 2.25\ i.e., N = 1.5$.

## 4.  Conclusion

In summary, we have numerically explored the propagation dynamics of chirped Airy pulse in a dispersive and highly nonlinear medium. The initial chirp proved to a crucial parameter to steer the dynamics of Airy pulses. Further, we have investigated the evolution of the maximum intensity point of the pulse (main pulse peak intensity) and its location to get a better understanding of the numerical results. We have found that, in general, the initial negative chirp increases the delays, and initial positive chirp reduces them in presence of SPM and SS in time domain. However, when the IRS being considered, the delay and the advancement depend on the magnitude of the initial chirp. In presence of SPM and SS both, a negative chirp enhances the blue-shifted component, and a positive chirp builds up a new red-shifted frequency component, as compared to the dynamics of an unchirped input pulse. Moreover, with IRS negative chirp broadens the spectrum significantly. Importantly, we have found that chirp can be used as an alternate way of frequency tuning and hence provides a new way to supercontinuum generation.

## Acknowledgements

Ankit Purohit thanks Ministry of Human Resource Development (MHRD), Government of India, for this work.

**Disclosures.** The authors declare no conflicts of interest.

**Data Availability.** Data underlying the results presented in this paper are not publicly available at this time but may be obtained from the authors upon reasonable request.